\documentclass{PoS}

\title{Long Range Topological Order, the Chiral Condensate, and the Berry Connection in QCD}

\ShortTitle{Long Range Topological Order in QCD}

\author{\speaker{H. B. Thacker}%
         \\
        University of Virginia\\
        E-mail: \email{hbt8r@virginia.edu}}

\abstract{
Topological insulators are substances which are bulk insulators but which carry current via special
``topologically protected'' edge states. The understanding of long range topological order in these systems is built around the idea
of a Berry connection, which is a gauge connection obtained from the phase of the electron wave function transported over
momentum space rather than coordinate space. The phase of a closed Wilson loop of the Berry connection around the Brillouin zone defines a topological
order parameter which labels discrete flux vacua. The conducting states are surface modes on the domain walls between discrete vacua.
Evidence from large-$N_c$ chiral dynamics, holographic QCD, and Monte Carlo observations has pointed to a picture of the QCD vacuum that
is very similar to that of a topological insulator, with discrete quasivacua labelled by $\theta$ angles that differ by mod $2\pi$.
In this picture, the domain walls are membranes of Chern-Simons charge, and the quark condensate consists of surface modes on these membranes,
which are delocalized and thus support the long range propagation of Goldstone pions. The Berry phase in QED2 describes charge polarization
of fermion-antifermion pairs, while in 4D QCD it describes the polarization of Chern-Simons membranes.
         }

\FullConference{32nd International Symposium on Lattice Field Theory \\
		 June 23 - 28 2014\\
		 Columbia University, NY}

\begin{document}

\section{Introduction}

The role of gauge field topology in resolving the $U_A(1)$ problem (i.e. the large mass of the flavor singlet $\eta'$ Goldstone boson) is well explored, both
theoretically and numerically. There is reason to believe that topology plays a much broader role in determining the low energy dynamics of QCD, 
but this role is less well understood. For the $\eta'$ problem, the quark-antiquark annihilation process that provides the gluonic component of the 
mass is highly localized and its contribution is determined simply by the topological susceptibility of the gluonic vacuum. Any model of the QCD vacuum that incorporates fluctuations
of topological charge density at the proper level (e.g. an instanton liquid model) will give the correct result, independent of the specific structure of the fluctuations. On the other hand, an understanding
of a possible topological origin of the chiral condensate and Goldstone pion propagation requires a theoretical framework that describes the long range structure of topological fluctuations. 
From the index theorem, we expect the near-zero eigenmodes of the Dirac operator that support the chiral condensate to be closely correlated with topological charge fluctuations
in the gauge field. For example, in the instanton liquid model the chiral condensate is formed from the 'tHooft zero modes of the instantons. However, it is difficult to understand
how massless pions propagate over long distances in an instanton framework. The problem is that instanton-like fluctuations are localized within approximately the confinement scale, 
so the quarks in the pion must hop from instanton to instanton. Unless the instantons are spatially correlated over long distances, they will not support the long range phase coherence required
for massless pion propagation. In the condensed matter analogy that I will explore in this talk, massless pion propagation is the analog of finite conductance, and its absence
in an uncorrelated instanton gas is an example of Anderson localization. More generally, it is not easy to reconcile the propagation of massless pions with 
the strong-coupling-inspired view of the QCD vacuum as a system in which the gauge fields are disordered over a distance larger than the confinement scale. The propagation
of the pion as a coherent chiral oscillation of the condensate would seem to require delocalized quark eigenmodes, suggesting a long range coherence in the topological
structure of the QCD vacuum. 

Early work by  Luscher \cite{Luscher78} and Witten \cite{Witten_largeN,Witten98} suggested the form of such long range fluctuations, namely, that they
are 2+1 dimensional membranes which are domain walls between discrete quasivacua, separating regions with values of the QCD vacuum $\theta$ angle differing by $\pm 2\pi$.
The $\theta$ angle serves as a topological order parameter which labels distinct quasivacua.
The existence of multiple discrete vacua separated by mod $2\pi$ steps in the $\theta$ angle is suggested by large-$N_c$ chiral Lagrangian arguments \cite{Witten_largeN}. Large $N_c$
phenomenology and the OZI rule indicate that the $U_A(1)$ anomaly appears in the chiral Lagrangian as a pure $\eta'$ mass term $\propto \eta'^2 = (-i\log {\rm Det} U)^2$ where
$U$ is the chiral field. The multiple branches of the log in this Lagrangian term imply multiple discrete, nearly degenerate quasivacua whose degeneracy is broken at $O(1/N_c)$ by the
$U_A(1)$ anomaly. The domain walls between these vacua are gauge excitations defined by the ``Wilson bag'' operator\cite{Luscher78}, the exponentiated integral of the 3-index Chern-Simons tensor 
integrated over the world volume of the membrane. The CS tensor is dual to the CS current, 
\begin{equation}
\label{eq:CScurrent}
K_{\mu} = \varepsilon_{\mu\alpha\beta\gamma}{\rm Tr}\left(A^{\alpha}\partial^{\beta}A^{\gamma} + \frac{2}{3}A^{\alpha}A^{\beta}A^{\gamma}\right)
\equiv 32\pi^2N_c\varepsilon_{\mu\alpha\beta\gamma}{\cal K}_3^{\alpha\beta\gamma}
\end{equation}
whose divergence is proportional to the topological charge density,
\begin{equation}
\label{eq:anomaly2}
\partial^{\mu}K_{\mu}={32\pi^2N_c}Q(x)
\end{equation}
From this it is easy to see that inserting a closed Chern-Simons membrane in the vacuum path integral is equivalent to including a $\theta$ term inside the enclosed 4-volume $V$, so
the CS membrane on surface $S$ is a domain wall separating vacua with distinct $\theta$ values, 
\begin{equation}
\label{eq:bag}
\exp \left(i\theta\int_V Q(x) d^4x\right) = \exp\left(i\theta\int_S {\cal K}_3^{\mu\nu\lambda}dx_{\mu}dx_{\nu}dx_{\lambda}\right)
\end{equation}
The value of $\theta/2\pi$ can be interpreted as the charge of the bag surface. The integral of ${\cal K}_3$ over the closed surface is not gauge invariant: it can change by an integer under
a topologically nontrivial gauge transformation on $S$. The right hand side of (\ref{eq:bag}) is thus only gauge invariant for bag charge $\theta/2\pi$ = integer.
Studies of the topological charge distribution in Monte Carlo gauge configurations \cite{Horvath03,Ilgenfritz07} have revealed long range structure in the form of thin, coherent
codimension one membranes of topological charge arranged in a layered array of alternating sign membranes. The spacing and thickness of the branes is of order the lattice spacing.
Oppositely charged surfaces are closely spaced, forming coherent dipole layers of topological charge.
The topological structure observed in Monte Carlo simulations is very naturally interpreted as a condensate of Chern-Simons membranes.

The observed structure of the TC distribution can actually be understood as a consequence of some basic, required features of the topological charge correlator.
The topological charge distribution is strongly constrained by the fact that the Euclidean correlator $G(x) = \langle Q(x)\;Q(0)\rangle$ 
{\it must be negative} for finite separation $|x|>0$. Furthermore, the integrated correlator is just the topological susceptibility, which is positive. The only way to accomodate 
both these features of the correlator is to have a positive ``wrong sign'' delta-function contact term at $x=0$. The observed layered, alternating sign array of coherent topological
charge membranes leads to exactly this form for the correlator in the continuum limit, with a dominant positive contact term at the origin and a negative tail beyond 2 or 3 
lattice spacings \cite{Horvath05}. The observed negativity of the correlator for $|x|>0$ beyond a few lattice spacings is a result of the alternating sign layered structure of the TC
distribution in the Monte Carlo configurations. (By contrast, a vacuum dominated by finite size instantons would not satisfy the negativity constraint for separations smaller than the instanton radius.)

In this talk I will discuss some of the implications of topological charge membranes for the chiral condensate and Goldstone boson propagation. 
General considerations surrounding the index theorem and spectral flow of the Dirac operator show that the topological charge membranes
will have attached Dirac eigenmodes, specifically, modes which are localized on the membrane surface but delocalized over its world volume. It is these topological surface modes which make up the
chiral condensate and provide for the long range propagation of massless pions. At this point we recognize a close parallel between this description of Goldstone boson propagation in the
QCD vacuum and recent developments in the theory of quantum conductivity in topological insulators and quantum Hall states. A topological insulator is a material which is a bulk insulator, i.e. the spectrum
for electrons propagating in the bulk has a mass gap, but which conducts electric current via massless, ``topologically protected'' boundary states which reside on domain wall 
surfaces. The surfaces separate regions with differing values of a topological order parameter. A central aspect of topological insulator theory is the Berry phase construction which
provides the definition of the topological order parameter. The Berry phase is obtained from the phase of a Bloch wave electron state under adiabatic transport in momentum space. For a particular
band, this defines a gauge connection over the Brillouin zone (BZ). A closed Wilson loop of the Berry connection which winds around the compact BZ is gauge
invariant under small topologically trivial transformations of the Berry connection. But the Wilson loop phase can change by integer multiples of $2\pi$ under large gauge
transformations, corresponding to threading units of ``magnetic'' flux through the loop that goes around the BZ. Such a discrete change of the Berry phase describes 
the transfer of a unit of charge between boundary surfaces. To formulate this idea we need to analytically
continue the phase of the wave function in momentum space, treating the Wilson loop as a closed contour in the complex momentum or rapidity plane, and explore the singularities of the analytically continued
Berry connection inside the loop. In general we expect the poles and cuts of analytically continued quark phase shifts in momentum space to reflect the low energy spectrum of the Dirac Hamiltonian,
so the nearby singularities of the analytically continued Berry phase are associated with the low-lying quark eigenmodes which form the condensate. 
In the case of QED2, it has been shown by studying the analytic structure of Bethe ansatz wave functions, 
that the poles in the Berry connection represent the vacuum polarization of quark-antiquark pairs \cite{Thacker-Wong}.

\section{Spectral flow, charge polarization, and the Berry connection}
To illustrate the connection between Hamiltonian spectral flow and the Berry phase description of topological charge, consider the example of 2D U(1) theory for the case of a constant
field strength on a periodic 2-torus with spatial length $2\pi$ and Euclidean time period $T$.
Choosing $A_0=0$ gauge, the gauge interaction term in the Dirac operator is $A_1 = Ft$, where $F\equiv \frac{1}{2}\epsilon^{\mu\nu}F_{\mu\nu}$ is the
field strength and $t$ is Euclidean time. Assume there is one unit of topological charge on the torus, so $F=1/T$. For a massless Dirac fermion in 2D Euclidean space, the Dirac eigenvalue equation separates into equations for left and right handed components:
\begin{eqnarray}
{\cal D}\psi_L(x,t)=\lambda \psi_L(x,t) \\
{\cal D}^*\psi_R(x,t)=\lambda \psi_R(x,t)
\end{eqnarray}
where
\begin{equation}
\label{eq:Dirac_op}
{\cal D} \equiv \frac{\partial}{\partial t} - i\frac{\partial}{\partial x} + Ft
\end{equation}
For $F=1/T>0$, there is one left-handed zero mode, given explicitly by
\begin{equation}
\label{eq:zeromode}
\psi_L(x,t=kT) = \sum_{n=-\infty}^{\infty} e^{-\frac{T}{2}(n+k)^2} e^{inx}
\end{equation}
which satisfies ${\cal D}\psi_L=0$.
If the field strength $F$ is negative, there is instead a right-handed zero mode $\psi_R = \psi_L^*$.

In $A_0=0$ gauge, the wavefunctions are quasiperidic in Euclidean time, i.e. periodic up to a gauge transformation,
\begin{equation}
\psi_L(x,t+T) = e^{-ix}\psi_L(x,t)
\end{equation}
In a Hamiltonian framework, the index theorem manifests itself in the form of a spectral flow constraint. Consider the Hamiltonian operator for a left-handed fermion 
as a parametric function of a rescaled time $k\equiv t/T$,
\begin{equation}
\label{eq:Hamiltonian}
H(t/T) = -i\frac{\partial}{\partial x} +\frac{t}{T}
\end{equation}
Acting on a periodic spatial interval $0<x<2\pi$, this Hamiltonian is periodic, up to a gauge transformation, over the Euclidean time interval $0<t<T$, i.e.
\begin{equation}
\label{eq:quasiperiodic}
H(k+1) = e^{-ix}H(k)e^{ix}
\end{equation}
Thus, for any integer $k$, the spectrum $E_n = n+k$ of $H(k)$ matches up with $E_n = n$, the spectrum of $H(0)$. The Dirac spectral flow for the gauge configuration is
given by $k$, the integer shift of the spectrum over the periodic time interval, or, more generally, the net number of left- minus right-handed modes that cross from negative
to positive energy. The spectral flow of the Dirac Hamiltonian is equal to the integrated topological charge over the 2D Euclidean space. 
The need for a gauge tranformation (\ref{eq:quasiperiodic}) to match wave functions at $t=0$ and $t=T$ is an indication that there has been a net transport of one unit of charge around 
the periodic spatial box $0<x<2\pi$. 
If we only consider the left chiral component of the Dirac field, the spectral evolution from $0$ to $T$ violates charge conservation, since an extra charge appears when an occupied
negative energy state crosses to positve energy. This is just the axial anomaly. For a full Dirac field, the left and right eigenstates mix as they cross $E=0$, and the spectral evolution
corresponds to pair production of a left-moving fermion and a right-moving antifermion.  (This can be shown by adding a small mass term to separate positive and negative energy branches
and then tracing the adiabatic evolution of the eigenstates \cite{Thacker14}.) Thus the spectral evolution induced by the background $F$ field conserves electric charge, but the
axial-vector charge in the box changes by +2 when the pair is created. .                                                            

Thus far, we have considered spectral flow in $A_0=0,\;A_1=Ft$ gauge, where $Ft\equiv k$ can be treated as a spectral parameter in the Hamiltonian. The relationship between spectral
flow and the Berry phase construction is seen by transforming to Coulomb gauge $A_0=-Fx,\; A_1=0$ with the gauge transformation 
\begin{equation}
g(x,t)=e^{iFtx} = e^{ikx}
\end{equation}
In this gauge, the Dirac Hamiltonian includes a Coulomb potential $-Fx$. 
The Hamiltonian eigenfunctions are now periodic in $t$ but quasiperiodic on the spatial interval $0<x<2\pi$. They have the form of Bloch wave states where the spectral parameter $k$ plays the role
of the Bloch wave momentum:
\begin{equation}
\Psi(x,k) = e^{ikx}u(x,k)
\end{equation}
where $u(x+2\pi,k) = u(x,k)$.
We can now interpret this as a Bloch wave defined on a large spatial volume $L\rightarrow\infty$ made up of unit lattice cells of length $2\pi$,
with a periodic coulomb potential..
Then an eigenstate on the periodic unit cell becomes a band of states on the Brillouin zone.
In this gauge, evolution over a Euclidian time period translates into adiabatic transport of the Bloch wave around the momentum space Brillouin zone.  
The Berry connection is obtained from the phase of the periodic part of the Bloch wave,
\begin{equation}
\label{eq:berryconn}
A(k) = {\rm Im} \int_0^{2\pi}dx\; u^*(x,k)\frac{\partial}{\partial k} u(x,k)
\end{equation}
The topological order parameter is the Berry phase given by the Wilson loop around the Brillouin zone
\begin{equation}
\label{eq:berryphase}
\theta = \oint dk A(k)
\end{equation}
This is the same construction that is central to the theory of electric charge polarization in topological insulators \cite{Vanderbilt}, and it has
the same physical interpretation. As I will discuss in the next section, the Berry phase for QED2 obtained in this way measures the electric polarization
of the vacuum due to fermion-antifermion pairs, while the analogous construction in 4D QCD measures the polarization of Chern-Simons membrane-antimembrane pairs in the gauge vacuum.   

\section{Pions as brane polarization waves}

To see the connection between the Berry phase (\ref{eq:berryphase}) and charge polarization in QED2, 
we introduce the bosonized form of the theory by writing the conserved vector current in terms of a real pseudoscalar field $\phi(x)$,
\begin{equation}
\label{eq:bosonization}
j_{\mu} = \frac{1}{2\pi}\varepsilon_{\mu\nu}\partial^{\nu}\phi
\end{equation}
Note that 2D bosonization (\ref{eq:bosonization}) is just the usual expression 
for electric charge polarization in terms of the corresponding currents, viz. $\vec{\nabla}\cdot\vec{P}=-j_0,\; \partial_0\vec{P}= \vec{j}$, 
with $\phi$ representing the polarization. a chiral rotation is a translation $\phi\rightarrow \phi+\alpha$. 
A fermion mass term corresponds to a sine Gordon interaction $\propto (1-cos\phi)$. In Coulomb gauge, the gauge 
interaction is a static Coulomb interaction which becomes a boson mass term $\propto e^2\phi^2$. 
The significance of the sine Gordon field for topological ordering is seen by considering the bosonic representation of fermions and
antifermions as topological kinks and antikinks where $\phi$ jumps by $\pm 2\pi$, it is easy to see that the pseudoscalar field is a topological order parameter whose vacuum value changes by $\pm2\pi$
when a pair is produced and the fermion and antifermion propagate to opposite boundaries of the cell. But this is just the process that results in a change of the Berry phase
(\ref{eq:berryphase}) by $\pm2\pi$. So we can identify the Berry phase on a cell with the average value of the polarization (i.e. the sine Gordon field) on the cell,
\begin{equation}
\label{eq:deltheta}
\Delta \theta = \Delta\int_0^{2\pi} \phi \frac{dx}{2\pi} = \int_0^{2\pi}\frac{dx}{2\pi}\int_0^Tdt\partial_0\phi = \int_0^Tdt\int_0^{2\pi}dx j_0^5 = \int_0^TQ_0^5dt
\end{equation}
The last expression shows that the Berry phase is a chiral phase rotation generated by the total axial charge on the unit cell. 

We can employ a similar Berry phase construction to describe the polarization of Chern-Simons membrane-antimembrane pairs in the QCD vacuum \cite{Thacker14}. 
The polarization of a brane pair is just the transverse separation between the brane and antibrane. 
In the case of QED2, the vacuum transition described by (\ref{eq:deltheta}) consists of a pair creation followed by
the propagation of the fermion and antifermion to opposite boundaries of the cell. The value of $\theta$ averaged over the cell at any time during this transition is just the spatial
separation of the pair, so when the fermion and antifermion reach opposite boundaries of the cell, we have $\theta=\pm 2\pi$. Similarly, in 4D QCD, the Berry 
phase describes the polarization of Chern-Simons membrane-antimembrane pairs. [Note that the fermion-antifermion pairs in QED2 can also be thought of as Chern-Simons membrane-antimembrane pairs, 
since in 2D a CS membrane is just a charged particle Wilson line.] With $\theta=0$ outside the bag formed by the membrane and antimembrane, we have $\theta=\pm2\pi$
inside the bag, and the value of $\theta$ averaged over the cell is just the transverse separation, i.e. the polarization. 

In a membrane condensate of the form described here, the distribution of topological charge transverse to the brane surfaces has an antiferromagnetic short range order at the scale of the cutoff,
as exhibited by the large negative 2-point TC correlator at a distance of a few lattice spacings. The integral of this correlator, which determines the topological susceptibility,
exhibits a large cancellation between the positive contact term at the origin and the short range negative contribution \cite{Horvath05}. The positive and negative contributions to $\chi_t$ are
separately divergent in the continuum limit, but the cancellation leads to a finite result for $\chi_t$ which scales nicely as the lattice spacing goes to zero. 
This provides good numerical evidence that the feature of the QCD vacuum that emerges in the continuum limit from the alternating sign ``membrane sandwich'' is its polarizability, 
which produces finite topological susceptibility of the gluonic vacuum. In this framework, a Goldstone pion is a bound state of a left-handed quark and a right-handed antiquark 
in surface modes of a membrane-antimembrane pair. Since discrete topological changes of the polarization $\theta$ correspond to $q\bar{q}$ pair creation or annihilation, we can
suppress these topological fluctuations by giving the quark and antiquark two different flavors. Then the polarization is described by a flavor nonsinglet polarization, obtained from
(\ref{eq:berryconn}) by putting flavor indices on the quark wave functions. In this case, the polarization of the chiral condensate represented by the Berry phase is generated
by the conserved nonsinglet axial charge. These nonsinglet polarization waves are Goldstone pions.

This work was supported by the Department of Energy under grant DE-SC00079984.

\end{document}